\documentclass[doublecol,linenumbers]{epl2} 
\usepackage{bm,multirow}
\usepackage{lipsum,capt-of,}
\usepackage{amssymb}
\usepackage{amsmath}

\usepackage{graphicx}
\newcommand{\jj}{\mathbf{j}}

\newcommand{\kk}{\mathbf{k}}
\newcommand{\vv}{\mathbf{v}}
\newcommand{\rr}{\mathbf{r}}
\newcommand{\uu}{\mathbf{u}}

\newcommand{\be}{\begin{equation}}
\newcommand{\ee}{\end{equation}}
\newcommand{\bea}{\begin{eqnarray}}
\newcommand{\eea}{\end{eqnarray}}
\newcommand{\ba}{\begin{align}}
\newcommand{\ea}{\end{align}}

\newcommand{\rmi}{{\rm i}}

\title{Superfluid liquid crystals: pasta phases in neutron star crusts}
\shorttitle{Superfluid liquid crystals} 

\author{D. N. Kobyakov\inst{1}  \and C. J. Pethick\inst{2,3}}
\shortauthor{D. N. Kobyakov and C. J. Pethick}

\institute{
  \inst{1} Institute of Applied Physics of the Russian Academy of Sciences, 603950 Nizhny Novgorod, Russia\\
\inst{2}The Niels Bohr International Academy, The Niels Bohr Institute, University of Copenhagen, Blegdamsvej 17,\\ DK-2100 Copenhagen \O, Denmark\\
 \inst{3}NORDITA, KTH Royal Institute of Technology and Stockholm University, Roslagstullsbacken 23,\\ SE-106 91 Stockholm, Sweden
}
\pacs{26.60.Gj}{Neutron star crust}
\pacs{47.37.+q}{Hydrodynamic aspects of superfluidity; quantum fluids}
\pacs{61.30.-v}{Liquid crystals}

\abstract{The pasta phases predicted to occur near the inner boundary of the crust of a neutron star resemble liquid crystals, a smectic A in the case of sheet-like nuclei (lasagna) and the columnar phase in the case of rod-like nuclei (spaghetti).  An important difference compared with usual liquid crystals is that the nucleons are superfluid.  We develop the hydrodynamic equations for this system and use them to study collective oscillations.  Nucleon superfluidity  leads to important qualitative differences in the spectra of these oscillations and also increases their frequencies compared with ordinary liquid crystals.  We discuss a number of directions for future work.}
\begin{document}

\maketitle

\section{Introduction}
In the inner crust of a neutron star it is predicted that, due to the competition between  the Coulomb energy and the nuclear surface  energy, nuclei will adopt shapes quite different from those of the roughly spherical nuclei found on Earth \cite{pasta}.   Because of their resemblance to the shapes of various varieties of pasta, these phases are often referred to as ``pasta" phases. These occur for densities slightly below that of the inner boundary of the crust of the star, about one half of the saturation density of nuclear matter with equal numbers of neutrons and protons. With increasing density, the nuclei become rod-like (spaghetti) and then  sheet-like (lasagna), and in both cases the nuclear matter is immersed in a neutron fluid and there is an essentially uniform background of electrons that ensures charge neutrality of bulk matter.  At yet higher densities, similar structures with the roles of the nuclear matter and neutron matter exchanged are a possibility but they are predicted not to be as prevalent as the spaghetti and lasagna phases.

To a first approximation, the spaghetti phase may be regarded as a periodic array of  columns of nuclear matter, while the lasagna phase is a periodic array of sheets of nuclear matter.    The spaghetti phase resembles that of a columnar liquid crystal, and the lasagna phase a smectic A liquid crystal, which in the following we shall refer to simply as ``smectic''.   However,  the physical origin of the structure is quite different, since in liquid crystals it is due to the shape of the constituent molecules.

In this Letter we develop a theory of the long wavelength dynamics, especially the collective modes, of the pasta phases.  Of particular interest here is the fact that the pasta can be distorted.  Our work may thus be regarded as an extension of work on collective oscillations in the pasta phases with the pasta assumed to be undistorted \cite{DiGalloOertelUrban}.
There are significant differences between the pasta phases and liquid crystals.  The first is that the pasta phases consist of neutrons, protons and electrons  and, consequently,  there is more than one mode involving changes in the particle densities, rather than the single one that exists in a one-component system.  The presence of the multiple components can be taken into account by a straightforward generalization of the standard theory.  The second difference is that the protons and neutrons in the pasta phases are superfluid.  Thus neutrons and protons can flow relative to the periodic structure of the pasta phases, while in liquid crystals, the mean velocity of the matter is equal to the velocity of the periodic structure to a good first approximation.  In the language of liquid crystals, deviations from this condition are referred to as ``permeation'' \cite{deGennesProst}   or ``percolation'' \cite{LandLElasticity}.  Since the neutrons and protons in the pasta phases are superfluid, permeation can play an important role, as we shall show.

In liquid crystals there are low-frequency modes whose velocities vanish for wave vectors parallel to the spaghetti strands or lying in the lasanga sheets.  This is easy to understand, since the restoring forces for such distortions of the structure are due to curvature terms in the energy, which vanish to leading order in the wave number.  What came as a surprise is that there are modes for which the velocity vanishes for wave vectors
perpendicular to the spaghetti strands or the lasagna sheets \cite{deGennes1969, MartinParodiPershan}, a prediction subsequently confirmed by experiment.  We show that, as a consequence of permeation, velocities of the corresponding modes of superfluid liquid crystals do not vanish.

 \section{Basic formalism}
  In neutron stars more than a year after their birth, temperatures have fallen to well below 10$^9$K ($\approx 0.1$MeV), so it is a good approximation to neglect the effect of thermal excitations  and the relevant thermodynamic potential is thus the energy.  The inner crust consists of neutrons, protons and electrons.  Our focus here is on low-frequency, long-wavelength phenomena and it is a
  good approximation to assume that net densities of electrical charge and electrical current are zero. Quantitatively, the conditions for this to hold for characteristic length scales $L$ and time scales $\cal T$ are, first, that $L\gg \lambda_D$, where $\lambda_D\approx ( \pi/ 4 \alpha)^{1/2}\hbar/p_{Fe}$ is the Debye screening length of the electrons and  $L/{\cal T}
 \ll v_e$, where $v_e$ is the electron velocity, which is close to the velocity of light, $c$.  Here $\alpha=e^2/\hbar c$ is the fine structure constant and $p_{Fe}$ is the electron Fermi momentum.   The collective modes to be considered have velocities considerably less than $c$, so the assumption of charge neutrality is valid for length scales large compared with the electron screening length, which is similar in order of magnitude but somewhat larger than the separation between pasta elements.  Under these conditions, the system may be regarded as having two constituents, the neutrons and the charged particles.

 Treating a superfluid using hydrodynamics is a good approximation provided that the characteristic length scale of disturbances is large compared with the coherence length of the nucleon superfluids, $\sim \hbar v/\pi \Delta$, $v$ being the nucleon Fermi velocity  and $\Delta$ the pairing gap.   Since the coherence length is typically comparable to the scale of the pasta structure, this condition does not place further restrictions on the validity of the hydrodynamic approach.

  To simplify the discussion, we shall first consider in detail the case of a single constituent, since this brings out the essential differences between ordinary liquid crystals and superfluid ones.  As we shall describe later, the effect of having more than one constituent does not change qualitatively the results for low-frequency modes associated with changes in the structure of the pasta, but it does lead to the existence of a second higher frequency density mode associated with counterflow.
 As in the case of liquid crystals \cite{deGennesProst}, we shall use as basic variables the local particle density, $n(\rr)$ and the displacement, $\uu(\rr)$, of a lasagna sheet perpendicular to the plane of the sheet or that of a spaghetti strand perpendicular to the strand.
\subsection{Kinetic energy}
 The kinetic energy in the pasta phases  depends on the velocity of the structure, $\partial\uu/\partial t$, and on properties of the superfluid.  To describe the latter, it is convenient to work in terms of the quantity $\phi$ which is one half of the phase of the pairing amplitude averaged over distances large compared with the scale of the structure but small compared with the scale $L$ of the phenomena under study; this was previously done to treat oscillations of those parts of the inner crust with roughly spherical nuclei \cite{CJPChamelReddy}.  By using this averaged phase, one filters out rapid variations of the phase on length scales comparable to the scale of the structure.  The momentum per particle in the condensate is $\hbar \bm\nabla\phi$.    First consider a situation where $\phi$ is independent of space.  The  total particle current density may be expressed phenomenologically as
  \be
  {\bf j}_0=n^n \left.\frac{\partial \uu}{\partial t}\right|_0,
  \ee
  where $n^n$ is the normal (particle number) density in the sense of a two-fluid model and the subscript ``0'' denotes that the derivative is to be evaluated in the reference frame in which $\phi$ is independent of space.  In general, $n^n$ is a function of $\partial \uu/\partial t$, but in this Letter we shall consider only linear phenomena, in which case the dependence of $n^n$ on $\partial \uu/\partial t$ may be ignored.

  The current density in a frame moving with velocity $-\bf V$ with respect to the first frame is given by the standard result for a Galilean transformation,
  \be
   {\bf j}=n^n \left.\frac{\partial \uu}{\partial t}\right|_0 + n {\bm V}.
   \ee
   In the new frame, $\uu=\uu_0+{\bf V}_\perp t$ and the phase is given by $\phi=m{\bf V}\cdot\rr/\hbar$, where $m$ is the particle mass.  The displacement $\uu$ is perpendicular to the lasagna sheets or spaghetti strands, so only  ${\bf V}_\perp$,  the component  of the velocity perpendicular to the sheets or strands enters the expression for the displacement in the second frame.  Thus the current density has the form
   \be
  {\bf j}  =n^n \frac{\partial \uu}{\partial t} + n^s \frac{\bm\nabla_\perp \phi}{m} +n \frac{\bm\nabla_\parallel \phi}{m}=  n^n \left(\frac{\partial \uu}{\partial t}-\frac{\bm\nabla_\perp \phi}{m}\right) +n \frac{\bm\nabla \phi}{m}.
   \label{current2}
  \ee
 In Eq. (\ref{current2}) the time derivative is to be evaluated in the frame moving with velocity $-\bm\nabla \phi/m$ with respect to the first frame and we now work in units in which $\hbar=1$.   The quantity $\bm\nabla_\parallel$ is the component of the gradient operator in the plane of the lasagna sheets or parallel to the spaghetti strands.   Equation (\ref{current2}) has the form expected for  a two-fluid model, but with an anisotropic normal density.  The quantity $n^s=n-n^n$ is the superfluid number density for currents perpendicular to the lasagna sheets or spaghetti strands, while for currents in the plane of the sheets or in directions parallel to the spaghetti strands, the normal density is zero, since we assume that the lasagna sheets are translationally invariant in those directions.

 In a similar fashion, one can calculate the kinetic energy density and one finds
 \bea
 E_{\rm kin} = \frac12 \frac{n}{m} (\bm\nabla_\parallel \phi)^2 +\frac12 \frac{n^s}{m} (\bm\nabla_\perp \phi)^2
 +\frac12 mn^n {\dot u}^2,
 \eea
 where the dot denotes the time derivative.
 This is similar to the expression in Ref. \cite{AndreevBashkin} for mixtures of quantum liquids except that here the normal fluid is due to the periodic structure, rather than thermal excitations.
\subsection{Potential energy}

    Here we shall treat only linear modes, and therefore we consider the deviation of the energy density $E=E_{\rm kin}+E_{\rm pot}$ from its value in the uniform state, with density $n_0$ and at rest, to second order in the density change $\delta n=n-n_0$, and the displacement $\uu$.  The potential energy density has the same form as for liquid crystals.  For lasagna it is given by the expression
  \begin{align}
 E_{\rm pot}(n, \uu)& =  E_{\rm pot}(n_0, 0)+\frac12 \frac{\partial^2E}{\partial n^2}(\delta n)^2 \nonumber\\+\frac{\partial^2E}{\partial n\partial (\bm\nabla_\perp\!\! \cdot{\bf u})}& \delta n \bm\nabla_\perp\!\! \cdot{\bf u} +\frac12 \frac{\partial^2E}{\partial (\bm\nabla_\perp \!\!\cdot{\bf u})^2} \left( \bm\nabla_\perp\!\! \cdot{\bf u}  \right)^2,\nonumber\\ &\hspace{10em} {\rm (lasagna)}.
 \end{align}
 To facilitate comparison with the results of Ref. \cite[\S 48]{LandLElasticity}, we introduce the notation
\be
E_{nn}=\frac{m}{n}A, \;\;\;  E_{n\bf u}=mC,\;\;\;{\rm and}\;\;\; E_{\bf u u}=nm B,
\label{ElastEnLasagna}
\ee
where the quantities $A$, $B$, and $C$, which have the dimensions of the square of a velocity, have the same meaning as in that reference.
Equation (\ref{ElastEnLasagna}) then becomes
 \bea
 E_{\rm pot}(n, \uu) = E_{\rm pot}(n_0, 0)+\frac12 \frac{mA}{n}(\delta n)^2 +mC \delta n \bm\nabla_\perp\!\! \cdot{\bf u}\nonumber\\ +\frac12 nmB \left( \bm\nabla_\perp\!\! \cdot{\bf u}  \right)^2, \,\,\,{\rm (lasagna)}. \,\,\,\,\,\,\,
 \eea
  For lasagna, the potential energy is relatively simple because the vector describing the distortions of the sheets has a single component, which is perpendicular to the sheets.   The only quantity required to define the state of the distorted lattice is the relative change in the separation of the sheets, $\bm\nabla_\perp\!\! \cdot{\bf u}$.  For spaghetti, the situation is more complicated because the vector describing the lattice distortion is two-dimensional, and two elastic constants are required, one to describe homologous distortions of the lattice, and another to describe shear.  If the strands lie in the  $z$-direction, the expression for the energy density  has the form  \cite[Eq. (7.30)]{deGennesProst}
 \begin{flalign}
  &E_{\rm pot}(n, \uu)= E_{\rm pot}(n_0, 0)+\frac12 \frac{mA}{n}(\delta n)^2 \nonumber\\&+mC \delta n \left(\frac{\partial u_x}{\partial x}+\frac{\partial u_y}{\partial y} \right)
 +\frac12 nmB_s \left( \frac{\partial u_x}{\partial x}+\frac{\partial u_y}{\partial y}  \right)^2\nonumber\\
 &+\frac12 nmB_t \left[\left( \frac{\partial u_x}{\partial y}+\frac{\partial u_y}{\partial x}        \right)^2+   \left( \frac{\partial u_x}{\partial x}-\frac{\partial u_y}{\partial y}        \right)^2 \right], \nonumber \\   &\hspace{15em}{\rm (spaghetti)},
  \label{El_energy_spaghetti}
 \end{flalign}
 where the subscript $s$ stands for ``scalar" and $t$ for ``tensor", since the components of the strain tensor in the two terms correspond to a two-dimensional scalar and a second rank tensor, respectively.
 \section{Mode velocities}
 To exhibit the effects of superfluidity we here calculate the mode velocities on the assumption that matter is incompressible, as de Gennes did for liquid crystals \cite{deGennes}.    The assumption of incompressibility is an excellent approximation for the low velocity modes associated with displacements of the structures since the elastic constant $A$, which is related to the bulk modulus of matter, is very much greater than $B$, $B_t$, and $C$, which depend on the Coulomb energy and the nuclear surface energy and are very much less than bulk nuclear energies.  We shall consider small-amplitude oscillations about a state in which matter is at rest, and therefore it is adequate to linearize the equations in the amplitude of the disturbance.
 The basic equations are the continuity equation expressing conservation of particle number,
   \be
\frac{\partial n}{\partial t} +\bm\nabla \cdot{\bf j}=0,
\label{cont_onecomp}
\ee
 and the condition that the rate of change of the momentum density, $m\bm j$, where $\bm j$ (Eq. (\ref{current2})) is equal to the force per unit volume, $\bm f$, or
 \be
   \frac{\partial {\bm j}}{\partial t}={\bm f}.
 \ee
 and $\bm f$ is given by
  \be
  {\bf f}=-\bm\nabla p-\frac{\delta E_{\rm pot}}{\delta \uu}.
  \label{force}
  \ee
   Here $p=n^2\partial (E/n)/\partial n$ is the pressure and the second term is due to elastic stresses in the structure. Changes in the pressure are given by $dp=nd\mu$, where $\mu=\partial E/\partial n$ is the chemical potential.  We shall also need an equation for the time development of the phase, which is a Josephson relation:
  \be
  \frac{\partial \delta \phi}{\partial t}=\mu-\mu_0= - \delta \mu.
  \label{Josephson}
  \ee
  Here $\delta \phi=\phi+\mu_0t$, where $\mu_0$ is the chemical potential in the unperturbed state.  Changes in the chemical potential are given by
  \be
  \delta \mu=\frac{\partial E_{\rm pot}}{\partial n}=\frac{mA}{n}\delta n +mC\bm\nabla_\perp\!\! \cdot{\bf u},
  \label{deltamu}
  \ee
  but for the calculations in this section, its precise form will not enter.
 \subsection{Lasagna}
  To be specific, we take the normal to the lasagna sheets to be in the $z$-direction and the currents within the plane to be in the $x$-direction.
For this choice of axes, only the $z$-component of the second term on the right side of Eq. (\ref{force}) is nonzero and it is given by
  \be
 -\frac{\delta E_{\rm pot}}{\delta u}= nmB\frac{\partial^2u}{\partial z^2} +mC\frac{\partial n}{\partial z}.
  \ee
  The equation of motion for the current density is

  \be
  \frac{\partial {\bm j}}{\partial t}=\frac{n}{m}\bm\nabla\dot\phi + n^{n}\left(\frac{\partial^2 \bm u}{\partial t^2} -\frac{{\bm\nabla}_\perp \dot \phi}{m}\right)=-\frac{n}{m}\bm\nabla \mu+nB{\bm\nabla}_{\perp}^2\bm u
  \label{eomj}
  \ee
  Here we have dropped the term involving $C$ since this involves density changes, which are neglected because of the assumption of incompressibility.
  From Eq. (\ref{cont_onecomp}), the incompressibility condition is  $\bm\nabla \cdot{\bf j}=0$, and, from Eqs. (\ref{eomj}), its time derivative thus gives
  \be
   \nabla^2 \delta \mu =mB\frac{\partial^3u}{\partial z^3} .\label{eomz}
  \ee
  If we take  the time and space dependence to be of the form $\exp\rmi(\kk\cdot\rr-\omega t)$, it follows from  the $z$-component of Eq. (\ref{eomj}) and Eq. (\ref{eomz}) that
  \be
  n^{n}\frac{\partial^2 u}{\partial t^2} = B\left(n-n^{n}\frac{k_z^2}{k^2}         \right)\frac{\partial^2u}{\partial z^2} ,
  \ee
  and the velocity of the mode, $c_2=\omega/k$ is given by
  \be
  c_2^2=  B\left(\frac{n}{n^n}-\cos^2\theta         \right)\cos^2 \theta,
  \label{c2lasagna}
  \ee
 where $\theta$ is the angle between the wave vector and the normal to the lasagna sheets.  We use the subscript ``2'' to denote this mode because of its similarity to second sound in liquid $^4$He \cite{deGennes}. Thus the velocity vanishes for $\theta=\pi/2$, as in the case of a smectic.
In a smectic, the normal density is equal to the total density and therefore the velocity is proportional to $|\sin\theta\cos\theta|$ and also vanishes for $\theta=0$.  However, in a superfluid liquid crystal ($n^s=n-n^n>0$), it does not.  This difference is due to the fact that in ordinary liquid crystals, permeation is neglected in the leading approximation.  Consequently, for $\theta=0$ the spatial modulation of the structure is locked to the matter and counterflow of the matter and the structure is impossible.  However, in superfluid liquid crystals, the modulations of the structure do not move with the matter, and an out-of-phase motion of the matter and the spatial structure is possible.

 The angular dependence of the velocity given by Eq. (\ref{c2lasagna}) may be expressed in  a more physical form by observing that the superfluid density tensor may be expressed as
\be
{\sf n}^s= n{\sf 1}-n^n {\hat {\bf z}}{\hat {\bf z}},
\ee
 where $\sf 1$ is the unit tensor and $\hat{\bf z}$ is a unit vector in the $z$-direction.  Therefore
 \be
 c_2^2=  B\frac{n^s(\theta)}{n^n}\cos^2 \theta,
 \ee
 where
 \be
 n^s(\theta) ={\hat \kk}\cdot {\sf n}^s\cdot{\hat \kk}
 \label{nstheta}
 \ee
 is the expectation value of the superfluid density tensor for a disturbance with wave vector $\kk$.  The factor ${n^s(\theta)}/{n^n}$ is familiar from the theory of second sound in liquid $^4$He \cite{Landau1941}, and the difference here is that the superfluid density depends on direction.  In addition, the factor $\cos^2\theta$ occurs because the forces on the sheets act only in the direction perpendicular to the sheets.

 Another way to understand the behavior of the mode velocity for small $\theta$ is to observe that in a smectic liquid crystal, a distortion of the sheets leads to large flow velocities in directions lying in the plane of the layers, since for incompressible matter with a constant density, $\bm\nabla\cdot\vv=0$ or $k_zv_z+k_xv_x=0$.  Here $\vv$ is the velocity of the matter.  Consequently $v_x=-v_z\cot \theta$ and for a given $v_z$, the transverse velocity diverges for $\theta\rightarrow 0$.  The kinetic energy density is given by
 \be
 E_{\rm kin}=\frac12mn(v_z^2+v_x^2)=\frac12 \frac{mn}{\sin^2\theta}v_z^2,
 \ee
 and thus one sees that the effective mass density associated with motion in the $z$-direction, $mn/\sin^2\theta$, diverges for $\theta\rightarrow 0$, and this makes the frequency of the mode vanish.  In a superfluid liquid crystal, counterflow of the structure and the superfluid is possible, and large flow velocities in directions lying in the plane of the lasanga sheets are not necessary to fulfil the incompressibility condition.

  \subsection{Spaghetti}

  For spaghetti, the calculations are simplified for specific choices of coordinates.  We take the spaghetti strands to lie parallel to the $z$-direction and the wave vector of the mode to lie in the $x$-$z$ plane.  With this choice, all derivatives with respect to $y$ vanish and  the energy density (\ref{El_energy_spaghetti}) then has the form
  \bea
  E_{\rm pot}(n, \uu)=E_{\rm pot}(n_0, 0)+\frac12 \frac{mA}{n}(\delta n)^2 +mC \delta n \frac{\partial u_x}{\partial x}\nonumber\\+\frac{nm}2 B \left( \frac{\partial u_x}{\partial x} \right)^2    +  \frac{nm}2 B_t \left( \frac{\partial u_y}{\partial x}\right)^2,
  \label{El_energy_spaghetti_xz}
  \eea
  where $B=B_s+B_t$.
  The momentum density is given by the current density, Eq. (\ref{current2}), times $m$ and its time rate of change is equal to the force per unit volume.  This gives
  \bea
  \frac{\partial \bm j}{\partial t}=\frac{n}{m}\frac{\partial \bm\nabla\phi}{\partial t} + n^{n}\left(\frac{\partial^2 \bm u}{\partial t^2} -\frac{1}{m}\frac{\partial \bm\nabla_\perp \phi}{\partial t}\right)\nonumber \\=-\frac{n}{m}{\bm\nabla \mu}- \frac{\delta E_{\rm pot}}{\delta \bm u} .
  \eea
For the choice of coordinates we have made, the components of the elastic contribution to the force per unit volume are given by
 \be
 -\frac{\delta E_{\rm pot}}{\delta u_x}=mnB\frac{\partial^2 u_x}{\partial x^2} \;\;\;{\rm and}\;\;\; -\frac{\delta E_{\rm pot}}{\delta u_y}= mnB_t \frac{\partial^2 u_y}{\partial x^2} .
 \ee
 In the first of these equations we have omitted the contribution proportional to $C$ because of the assumption of incompressibility.
 The time derivative of the incompressibility condition, $\partial(\bm\nabla \cdot {\bf j})/\partial t =0$, is therefore
\be
\frac{n}{m}\nabla^2 \mu=B\frac{\partial^3u_x}{\partial x^3}.
\label{incomp_spag}
\ee
The equations of motion for the displacements are
 \be
 mn^{n}\frac{\partial^2u_x}{\partial t^2}=-n^{n}\frac{\partial \mu}{\partial x}+B\frac{\partial^2u_x}{\partial x^2}
 \ee
 and
 \be
  mn^{n}\frac{\partial^2u_y}{\partial t^2}=B_t\frac{\partial^2u_y}{\partial x^2}.
\ee
 On using the incompressibility condition (\ref{incomp_spag}) to eliminate $\mu$, one finds that there is one mode in which only $u_x$ is nonzero and it has a velocity $c_2$ given by
 \be
 c_2^2=B\frac{n}{n^{n}}\sin^2 \theta \left(1-\frac{n^{n}}{n}\sin^2 \theta \right),
 \ee
 where $\sin \theta=k_x/k$.  This result is similar to that for the mode in lasagna, and the fact that $\sin^2\theta$ occurs here rather than $\cos^2 \theta$ reflects the fact that for lasagna the superfluid density is reduced for motion in the $z$-direction, whereas for spaghetti the reduction occurs for motion perpendicular to the $z$-direction.

 The results for spaghetti may be expressed in terms of an angle-dependent superfluid density, as was done for lasagna.  For spaghetti, the expression for the superfluid density tensor is
\be
{\sf n}^s= n {\sf 1}-n^n ({\hat {\bf x}}{\hat {\bf x}}+{\hat {\bf y}}{\hat {\bf y}}),
\label{ns_spaghetti}
\ee
 where $\hat{\bf x}$  and  $\hat{\bf y}$ are unit vectors in the $x$- and $y$-directions.  Thus
 \be
 c_2^2=  B\frac{n^s(\theta)}{n^n}\sin^2 \theta,
 \ee
 where $n^s(\theta)$ is calculated from Eq. (\ref{nstheta}) but with the superfluid density tensor appropriate for spaghetti, Eq. (\ref{ns_spaghetti}).

  The other mode is transverse, with displacements only in the $y$-direction, and it has a velocity $c_t$ given by
 \be
  c_t^2=\frac{B_t}{mn^{n}}\sin^2 \theta.
  \label{c_t}
 \ee
 In this mode, which is a new feature of spaghetti compared with lasagna,  the displacement is perpendicular to both the direction of the strands and to the wave vector.  Thus it is a purely transverse wave, which accounts for our denoting its velocity by $c_t$, and it reduces to the result for columnar liquid crystals \cite[Sec. 8.1.7]{deGennesProst} when $n^n=n$.

 \section{Generalization to two constituents}
 We now generalize the discussion to the case of two constituents, neutrons and charged particles.  The density of neutrons is denoted by $n_n$ and that of protons by $n_p$.  Because matter is electrically neutral, the electron density is equal to $n_p$. We denote the phases of the neutron and proton condensates by $2\phi_n$ and $2\phi_p$, respectively.  For simplicity, we shall neglect the difference between the proton and neutron masses because this is small compared with the effective mass of an electron, $\approx p_{Fe}/c$, and relativistic contributions to the masses of the nucleons, which we also neglect.  The basic ideas described earlier for a single constituent can be used, but the expressions for the current densities are more complicated because of the entrainment of neutrons by protons, and vice versa.   The natural generalization of Eq. (\ref{current2}) is
 \bea
  {\bf j} _n =n_n^n \frac{\partial \uu}{\partial t} + n_{nn}^{s\perp} \frac{\bm\nabla_\perp \phi_n}{m} + n_{np}^{s\perp} \frac{\bm\nabla_\perp \phi_p}{m}\nonumber \\+n_{nn}^{s\parallel} \frac{\bm\nabla_\parallel \phi_n}{m} +n_{np}^{s\parallel} \frac{\bm\nabla_\parallel \phi_p}{m},
   \label{current2n}
  \eea
 and
  \bea
  {\bf j} _p =n_p^n \frac{\partial \uu}{\partial t} + n_{pp}^{s\perp} \frac{\bm\nabla_\perp \phi_p}{m} + n_{np}^{s\perp} \frac{\bm\nabla_\perp \phi_n}{m}\nonumber \\+n_{pp}^{s\parallel} \frac{\bm\nabla_\parallel \phi_p}{m} +n_{np}^{s\parallel} \frac{\bm\nabla_\parallel \phi_n}{m}.
   \label{current2n}
  \eea
  Here the superfluid number density tensor
  is defined by
  \be
  n^\perp_{\alpha\beta}=m\frac{\partial^2 E}{\partial ({\bm\nabla}_\perp \phi_\alpha) \partial ({\bm\nabla}_\perp \phi_\beta)}
  \ee
  and the corresponding equation for $n^\parallel_{\alpha\beta}$.
  From Galilean invariance it follows that
  \bea
   n_n^n  + n_{nn}^{s\perp} + n_{np}^{s\perp}=n_{nn}^{s\parallel}  +n_{np}^{s\parallel}=n_n
  \eea
  and
  \bea
  n_p^n  + n_{pp}^{s\perp} + n_{np}^{s\perp}=n_{pp}^{s\parallel}  +n_{np}^{s\parallel}=n_p.
  \eea

 The total nucleon number density is $n=n_n+n_p$, the total current density is $\jj=\jj_n+\jj_p$ and the momentum density is $m\bf j$.  The rate of change of the momentum density is given by the force per unit volume, Eq. (\ref{force}), and combining this with the equation of continuity, as was done for the one-component case, one arrives at Eq. (\ref{c2lasagna}), where the total normal density is now $n^n=n_n^n+n_p^n$.

\section{Coupling to density variations}
In the calculations above, we have taken the matter to be incompressible.  When this assumption is relaxed, modes associated with oscillations of the structure of the pasta are coupled to sound waves.  To illustrate these effects, we consider the case of lasagna and restrict ourselves to one constituent.

The current density satisfies the equations
\be
\frac{\partial j_z}{\partial t}=n^n \frac{\partial^2 u}{\partial t^2} +n^s\frac{\partial^2 \phi}{\partial t\partial z}=-\frac{1}{m}\frac{\partial p}{\partial z}+n B \frac{\partial^2u}{\partial z^2}+C\frac{\partial n}{\partial z}
\label{eom_gz}
\ee
and
\be
\frac{\partial j_x}{\partial t}=\frac{n}{m}\frac{\partial^2 \phi}{\partial t\partial x}=-\frac{1}{m}\frac{\partial p}{\partial x}.
\label{eom_gx}
\ee
We eliminate the pressure changes and the phases of the condensate from Eq. (\ref{eom_gz})   by using the fact that
\be
\delta p=n\delta \mu=Am\delta n +mn C\frac{\partial u}{\partial z}
\label{deltap}
\ee
and that the phase satisfies the Josephson relation (\ref{Josephson}).
One then finds
\be
n^n \frac{\partial^2 u }{\partial t^2}=   ( nB-n^n C)\frac{\partial^2 u}{\partial z^2} +  \left( C -\frac{n^n}{n}A\right)\frac{\partial n}{\partial z},
\ee
or, when Fourier transformed, with  the mode velocity defined by $c=\omega/k$,
\be
\left[c^2-\left(\frac{n}{n^n} B- C\right)\cos^2\theta\right]u=-\rmi \frac{\cos \theta}{k}\left( \frac{C}{n^n}-\frac{A}{n}   \right)\delta n.
\label{rho_u1}
\ee
The time derivative of the continuity equation (\ref{cont_onecomp}) is
\be
\frac{\partial^2 n}{\partial t^2}-\left(\frac{\partial^2}{\partial x^2} + \frac{\partial^2}{\partial z^2}   \right)\frac{p}{m}+n B \frac{\partial^3u}{\partial z^3} +C\frac{\partial^2 n}{\partial z^2}=0.
\label{t_deriv_cont}
\ee

Substituting Eq. (\ref{deltap}) into Eq. (\ref{t_deriv_cont}) and Fourier transforming,  one finds
\be
(c^2-A+C\cos^2\theta)\delta n=\rmi k n\cos\theta  (C  -B\cos^2\theta)u .
\label{rho_u2}
\ee
From Eqs. (\ref{rho_u1}) and (\ref{rho_u2}) one finds the following equation for the mode velocities:
\bea
c^4-c^2\left(A+\left[\frac{n}{n^n}B-2C \right]\cos^2\theta      \right)\nonumber \\+(AB-C^2)\cos^2\theta\left( \frac{n}{n^n} -\cos^2\theta  \right) =0.
\label{Lasagnamodes_comp}
\eea
When the superfluid density is zero, $n^n=n$ and the result reduces to that of Ref. \cite[p.184]{LandLElasticity}.

In the pasta phases, the quantities $B$ and $C$, which depend on Coulomb and nuclear surface energies are much less in magnitude than $A$, which is related to bulk energies of nuclear and neutron matter.  In this limit, the mode associated with lattice distortions has a velocity given by Eq. (\ref{c2lasagna}) while the density mode has a velocity  given by $c_1^2=A$.  This is to be contrasted with the case of rigid pasta, which corresponds to taking in Eq. (\ref{Lasagnamodes_comp}) the limit $B\rightarrow \infty$ with $A$ and $C$ finite.  One finds $c_1^2 \simeq A n^s(\theta)/n$:  the velocity depends on angle and is lower than that for flexible pasta.  The physical point is that, in the pasta phases, the structure is relatively easily distorted, and in the density modes it is ``swept along'' with the superfluid.  Consequently, in these modes both normal and superfluid components move.

In this section, we have assumed that there is a single constituent.  To take into account the fact that the pasta phases have two superfluid constituents, neutrons and protons, one may generalize the discussion to that case with nothing new conceptually but at the expense of algebraic complexity.  The qualitatively new feature is that there are two density modes, one in which the neutron and proton densities are in phase and another in which they are out of phase.  This is analogous to the two modes found in the uniform liquid phase of neutron star matter, as discussed in Refs. \cite{Two_component}.
The discussion for spaghetti follows closely that for lasagna.  As we have seen, in addition to the two modes found for lasagna, there is a transverse one due to the nonzero shear rigidity of the lattice.  Since the latter mode does not couple to density oscillations, its velocity is given by the result obtained assuming matter to be incompressible, Eq.\,(\ref{c_t}).  The two other modes may be found by generalizing the discussion of the previous section to spaghetti.   The dispersion relation is the same as Eq. (\ref{Lasagnamodes_comp})
 for lasagna, but with $\cos\theta$ replaced by $\sin\theta$.

\section{Dispersion}
 In order to calculate quantities such as the heat capacity, which depend on the detailed behavior of the collective mode spectrum for small $\bf k$, it is necessary to take into account  terms of higher order in the gradient expansion.   For lasagna, the leading such term in the energy density is
\be
\Delta E=\frac12 K_1\left(\nabla_{\parallel}^2u\right)^2=\frac12 K_1 \left( \frac{\partial^2u}{\partial x^2} +\frac{\partial^2u}{\partial y^2}   \right)^2,
\ee
and for spaghetti it is
\be
\Delta E=\frac12 K_3\left(\nabla_{\parallel}^2{\bf u}\right)^2=\frac12 K_3 \left[ \left(\frac{\partial^2u_x}{\partial z^2}\right)^2+\left(\frac{\partial^2u_y}{\partial z^2}\right)^2\right],
\ee
where the second expressions in these equations apply for the specific choice of axes we have made.  The quantities $K_1$ and $K_3$ are elastic constants in the notation of Ref. \cite{CJPPotekhin}.   When these terms are included in the calculation of mode frequencies, $\omega_\kk$, one finds for second sound  when the matter is treated as incompressible
the expressions
\be
\omega_2(\kk)^2= {B(k_z^2+k_\perp^4\ell_1^2)}\left(\frac{n}{n^n} -\frac{k_z^2}{k^2}\right),
 \,\,\,{\rm (lasagna)},
 \label{omega2lasagna}
\ee
and
 \be
\omega_2(\kk)^2= {B(k_\perp^2+k_z^4\ell_3^2)}\left(\frac{n}{n^n} -\frac{k_\perp^2}{k^2}\right),
 \,\,\,{\rm (spaghetti)},
\ee
where $\ell_1=(K_1/nmB)^{1/2}$ and $\ell_3=(K_3/nmB)^{1/2}$.
For the transverse mode in spaghetti the result is
 \be
\omega_t(\kk)^2= {B_t(k_\perp^2+k_z^4\ell_t^2)}\left(\frac{n}{n^n} -\frac{k_\perp^2}{k^2}\right),
 \,\,\,{\rm (spaghetti)},
\ee
where $\ell_t=(K_3/nmB_t)^{1/2}$.  As we shall show in future work \cite{KobyakovCJP}, the heat capacity due to collective modes varies as $T^2$ for lasagna and $T^{5/2}$ for spaghetti at low temperatures, and it is therefore larger than for ordinary phonons, for which it varies as $T^3$.

\section{Discussion and concluding remarks}
In this Letter we have derived hydrodynamic equations to describe the dynamics of the pasta phases.  The model is in essence a three-fluid one, with a ``normal'' velocity associated with motion of the structure, and two superfluid momenta associated with the neutron and proton condensates.   A key difference between the pasta phases and ordinary liquid crystals is that the structure does not move with the local velocity of the matter.   We find that oscillation frequencies of modes involving changes in the structure are higher than they would be if the nucleons were normal.

As has been demonstrated for smectic liquid crystals \cite{Caille} and for pion condensates in nuclear matter \cite{BaymFrimanGrinstein}, there is no long range order in infinite systems at nonzero temperature.
 This also applies for the lasagna phase, but further work is required to determine how important the effect is quantitatively.

 To make numerical calculations of mode frequencies, one requires values for the elastic constants and of the components of the superfluid density tensor.  The elastic constants $B$ and $B_t$ have been calculated in Ref. \cite{CJPPotekhin} and estimates of the neutron normal density for the pasta phases may be obtained from the work of Ref. \cite{WatanabeCJP}.  These calculations need to be extended to obtain an improved understanding of mode frequencies.

In our calculations, we have worked in the harmonic approximation and have not taken into account nonlinear coupling of modes, which leads to anomalous behavior of the elastic properties at very long wavelengths and low frequencies \cite{MartinParodiPershan, GrinsteinPelcovits}.  In ordinary liquid crystals, such effects are difficult to observe experimentally but the question of whether or not they can be important in the pasta phases remains open.

In the calculations presented here, it has been assumed that the lasagna sheets and spaghetti strands are uniform, whereas microscopic studies indicate that  at low temperature the thickness of the sheets and the transverse area of the strands are modulated in space \cite{WilliamsKoonin,PaisandOthers}.   However, at some nonzero temperature one expects the modulation in the plane of the sheets or along the spaghetti strands to ``melt'', and our treatment will be valid above this temperature.  At lower temperatures, to describe distortions of the structure, it is necessary to use a three dimensional vector $\bm u$, since displacements of the structure lying in the plane of the sheets or along the direction of the strands will affect the energy.  As a consequence, the oscillations of the pasta phases are expected to have frequencies linear in the wave vector in all directions.

Throughout, we have assumed that the pasta is perfectly ordered, whereas in practice it is highly likely that there will be imperfections such as those that have been found in numerical simulations \cite{Horowitz}.  There could also be regions with more complicated phases that have been found in Hartree--Fock \cite{PaisandOthers}, time-dependent Hartree--Fock \cite{Sebille} and molecular dynamics {\cite{Sonoda} calculations.  Quite generally, the pasta phases could have complicated textures.  Important problems for future work are to determine the density of imperfections, and the spatial scale of the textures.  These affect the mechanical properties of the pasta phases, including the elastic constants and yield stresses, which are significant quantities for estimating the strength of gravitational wave emission from  rotating neutron stars.

\acknowledgments
We are grateful to Gerd Schr\"oder-Turk for informative discussions and Lev Pitaevskii for valuable correspondence.
This work was supported by the Russian Foundation for Basic Research, under research project No. 31 16-32-60023 mol$_-$a$_-$dk.

\end{document}